 \documentclass[final,5p,times,twocolumn]{elsarticle}

\usepackage{graphicx}
\usepackage{amssymb}
\usepackage[tbtags]{amsmath}
\biboptions{sort&compress}

\renewcommand{\epsilon}{\varepsilon}

\journal{Physics Letters B}

\begin{document}

\begin{frontmatter}



%\tnotetext[SUPP]{
%Partially supported by 
%the Russian Foundation for Basic Research, grants 08-02-00258, 08-02-00258, and RF 
%Presidential Grant for Sc. Sch. NSh-5655.2008.2.
%}

\title{Measurement of $J/\psi\to\gamma\eta_{\rm c}$ decay rate 
and $\eta_{\rm c}$ parameters at KEDR
%\tnoteref{SUPP}
}


\author[ad1]{\mbox{V.V. Anashin}}
\author[ad1,ad2]{\mbox{V.M. Aulchenko}}
\author[ad1,ad2]{\mbox{E.M. Baldin}}
\author[ad1]{\mbox{A.K. Barladyan}}
\author[ad1,ad2]{\mbox{A.Yu. Barnyakov}}
\author[ad1,ad2]{\mbox{M.Yu. Barnyakov}}
\author[ad1,ad2]{\mbox{S.E. Baru}}
\author[ad1]{\mbox{I.Yu. Basok}}
\author[ad1]{\mbox{I.V. Bedny}}
\author[ad1,ad2]{\mbox{A.E. Blinov}}
\author[ad1,ad2,ad3]{\mbox{V.E. Blinov}}
\author[ad1]{\mbox{A.V. Bobrov}}
\author[ad1]{\mbox{V.S. Bobrovnikov}}
\author[ad1,ad2]{\mbox{A.V. Bogomyagkov}}
\author[ad1,ad2]{\mbox{A.E. Bondar}}
\author[ad1]{\mbox{A.R. Buzykaev}}
\author[ad1,ad2]{\mbox{S.I. Eidelman}\corref{cor}}
\ead{S.I.Eidelman@inp.nsk.su}
\author[ad1]{\mbox{Yu.M. Glukhovchenko}}
\author[ad1]{\mbox{V.V. Gulevich}}
\author[ad1]{\mbox{D.V. Gusev}}
\author[ad1]{\mbox{S.E. Karnaev}}
\author[ad1]{\mbox{G.V. Karpov}}
\author[ad1]{\mbox{S.V. Karpov}}
\author[ad1,ad2]{\mbox{ T.A. Kharlamova}}
\author[ad1]{\mbox{V.A. Kiselev}}
\author[ad1,ad2]{\mbox{S.A. Kononov}}
\author[ad1]{\mbox{K.Yu. Kotov}}
\author[ad1,ad2]{\mbox{E.A. Kravchenko}}
\author[ad1,ad2]{\mbox{V.F. Kulikov}}
\author[ad1,ad3]{\mbox{G.Ya. Kurkin}}
\author[ad1,ad2]{\mbox{E.A. Kuper}}
\author[ad1,ad3]{\mbox{E.B. Levichev}}
\author[ad1]{\mbox{D.A. Maksimov}}
\author[ad1]{\mbox{V.M. Malyshev}\corref{cor}}
\ead{V.M.Malyshev@inp.nsk.su}
\author[ad1]{\mbox{A.L. Maslennikov}}
\author[ad1,ad2]{\mbox{A.S. Medvedko}}
\author[ad1,ad2]{\mbox{O.I. Meshkov}}
\author[ad1]{\mbox{S.I. Mishnev}}
\author[ad1,ad2]{\mbox{I.I. Morozov}}
\author[ad1,ad2]{\mbox{N.Yu. Muchnoi}}
\author[ad1]{\mbox{V.V. Neufeld}}
\author[ad1]{\mbox{S.A. Nikitin}}
\author[ad1,ad2]{\mbox{I.B. Nikolaev}}
\author[ad1]{\mbox{I.N. Okunev}}
\author[ad1,ad3]{\mbox{A.P. Onuchin}}
\author[ad1]{\mbox{S.B. Oreshkin}}
\author[ad1,ad2]{\mbox{I.O. Orlov}}
\author[ad1]{\mbox{A.A. Osipov}}
\author[ad1]{\mbox{S.V. Peleganchuk}}
\author[ad1,ad3]{\mbox{S.G. Pivovarov}}
\author[ad1]{\mbox{ P.A. Piminov}}
\author[ad1]{\mbox{V.V. Petrov}}
\author[ad1]{\mbox{A.O. Poluektov}}
\author[ad1]{\mbox{G.E. Pospelov}}
\author[ad1]{\mbox{V.G. Prisekin}}
\author[ad1,ad2]{\mbox{O.L. Rezanova	}}
\author[ad1]{\mbox{A.A. Ruban}}
\author[ad1]{\mbox{V.K. Sandyrev}}
\author[ad1]{\mbox{G.A. Savinov}}
\author[ad1]{\mbox{A.G. Shamov}}
\author[ad1]{\mbox{D.N. Shatilov}}
\author[ad1,ad2]{\mbox{B.A. Shwartz}}
\author[ad1]{\mbox{E.A. Simonov}}
\author[ad1]{\mbox{S.V. Sinyatkin}}
\author[ad1]{\mbox{A.N. Skrinsky}}
\author[ad1,ad2]{\mbox{V.V. Smaluk}}
\author[ad1]{\mbox{A.V. Sokolov}}
\author[ad1]{\mbox{A.M. Sukharev}}
\author[ad1,ad2]{\mbox{E.V. Starostina}}
\author[ad1,ad2]{\mbox{A.A. Talyshev}}
\author[ad1]{\mbox{V.A. Tayursky}}
\author[ad1,ad2]{\mbox{V.I. Telnov}}
\author[ad1,ad2]{\mbox{Yu.A. Tikhonov}}
\author[ad1,ad2]{\mbox{K.Yu. Todyshev}}
\author[ad1]{\mbox{G.M. Tumaikin}}
\author[ad1]{\mbox{Yu.V. Usov}}
\author[ad1]{\mbox{A.I. Vorobiov}}
\author[ad1]{\mbox{A.N. Yushkov}}
\author[ad1]{\mbox{V.N. Zhilich}}
\author[ad1,ad2]{\mbox{V.V.Zhulanov}}
\author[ad1,ad2]{\mbox{A.N. Zhuravlev}} 

\cortext[cor]{Corresponding authors}

\address[ad1]{Budker Institute of Nuclear Physics, 11, Lavrentiev prospect, Novosibirsk, 630090, Russia}
\address[ad2]{Novosibirsk State University, 2, Pirogova street, Novosibirsk, 630090, Russia}
\address[ad3]{Novosibirsk State Technical University, 20, Karl Marx prospect, Novosibirsk, 630092, Russia}

\begin{abstract}
Using the inclusive photon spectrum
based on a data sample collected at the $J/\psi$ peak
with the KEDR detector at the VEPP-4M $e^+e^-$ collider, 
we measured the rate 
of the radiative decay $J/\psi\to\gamma\eta_{\rm c}$ as well as $\eta_{\rm c}$ 
mass and width. Taking into account an asymmetric  
photon lineshape we obtained
$\Gamma^0_{\gamma\eta_{\rm c}}=2.98\pm0.18 \phantom{|}^{+0.15}_{-0.33}$ keV,
$M_{\eta_{\rm c}} = 2983.5 \pm 1.4 \phantom{|}^{+1.6}_{-3.6}$ MeV/$c^2$, 
$\Gamma_{\eta_{\rm c}} = 27.2 \pm 3.1 \phantom{|}^{+5.4}_{-2.6}$ MeV. 
\end{abstract}


\end{frontmatter}

% \linenumbers


\section{Introduction}	
%%\label{}

$J/\psi\to\gamma\eta_{\rm c}$ decay is a magnetic dipole radiative transition
in charmonium with the most probable photon energy $\omega_0$ of about 114 MeV
and a fairly large branching fraction of $(1.7\pm0.4)\%$~\cite{PDG}.
This is a transition between 1$S$ states of the charmonium system 
and its rate can be easily calculated in potential models. 
In the nonrelativistic approximation, the magnetic dipole amplitudes 
between $S$-wave states are independent of a
specific potential model, because the spatial overlap equals one for states 
within the same multiplet.
A simple calculation in the nonrelativistic approximation yields the 
result~\cite{EICH} ${\cal B}(J/\psi\to\gamma\eta_{\rm c})=3.05\%$.
It is reasonable to assume that relativistic corrections are of order $20\div30$\%, 
similarly to the case of the electric dipole transitions 
in the charmonium (see, for example, the reviews~\cite{QWG,strong}). 
However, in 1986 the   
Crystal Ball Collaboration measured this branching fraction 
in the inclusive photon spectrum and obtained a much smaller value 
$(1.27\pm0.36)\%$~\cite{GAIS}. 
There are a lot of theoretical predictions 
for this decay rate~\cite{SHIF,KHOD,BEIL,BRAM,DUDE,DONALD,Becirevic,Pineda}, 
based on QCD sum rules, 
lattice QCD calculations and so on, but as a rule they lead to values
 approximately twice as large as the Crystal Ball result.

This discrepancy remained unchanged for more than twenty years. 
During this period no new measurements of 
this branching fraction were performed, and the PDG average~\cite{PDGold} 
was based on the single Crystal Ball result. Only in 2009 the CLEO
Collaboration published the result of a new 
measurement~\cite{MITC}, in which 
12 exclusive decay modes of the $\eta_{\rm c}$ were analyzed. The obtained value
${\cal B}(J/\psi\to\gamma\eta_{\rm c})=(1.98\pm0.09\pm0.30)\%$ is 
closer to theoretical predictions. Combining the Crystal Ball and 
CLEO results, PDG obtained 
${\cal B}(J/\psi\to\gamma\eta_{\rm c})=(1.7\pm0.4)\%$~\cite{PDG} with a scale factor 
of 1.6. In this work we report the result of a new independent measurement
performed using the inclusive photon spectrum.

\section{Photon spectrum}
The spectrum of detected photons in 
$J/\psi \to \gamma\eta_{\rm c}$ decay is given by the formula~\cite{QWG}
\begin{equation}
\label{fone}
\frac{d\Gamma(\omega)}{d\omega} = \frac{4}{3}\alpha \frac{e_{\rm c}^2}{m_{\rm c}^2}\omega^3|M|^2\textnormal{BW}(\omega).
\end{equation}
Here $\omega$ is a photon energy, $\alpha$ is the fine structure constant, $e_{\rm c}$ and $m_{\rm c}$ are $c$-quark charge 
(in electron charge units) and mass, $M=\left<\eta_{\rm c}|j_0(\omega r/2)|J/\psi\right>$ is the matrix element of 
the transition (without relativistic corrections), $j_0(x)=\sin(x)/x$, $\textnormal{BW}(\omega)$ is a Breit-Wigner function.
A typical momentum transfer inside the charmonium bound state is about 700 to 800 MeV~\cite{BRV} (this is of
the order of the inverse size of the system), 
so the matrix element is almost constant (close to one) up to such photon energies. 
Therefore, in this energy range the decay spectrum $\frac{d\Gamma(\omega)}{d\omega} \sim \omega^3 \textnormal{BW}(\omega)$. 
Since $\textnormal{BW}(\omega)\sim\omega^{-2}$ at $\omega \gg \omega_0$, the decay probability grows as $\omega$ when $\omega$ increases. 
If a resonance width is not small, it can give a noticeable tail in the photon spectrum at photon energies $\omega\gg\omega_0$.
For the $J/\psi\to\gamma\eta_{\rm c}$ transition we have 
$\frac{\Gamma_{\eta_{\rm c}}}{\omega_0}\approx\frac{30~\textnormal{MeV}}{114~\textnormal{MeV}}\approx\frac{1}{4}$.
This value is not small, therefore we should take into account this tail. 
It should be also noted 
that in theoretical calculations of the decay rate this effect is
 as a rule neglected, and, assuming a small width of the resonance,  
$\omega$ is replaced with $\omega_0$ in (\ref{fone}).

At the same time it is known that the usual form of the Breit-Wigner function is applicable only in the close vicinity of a resonance
and gives an overestimated value far from it. For example, in the theory 
of atomic transitions a photon absorption lineshape has the 
same functional form as a (non-relativistic) Breit-Wigner function, but with $\Gamma(\omega)\sim \omega^3$~\cite{LOUD}, 
so $\textnormal{BW}(\omega)\sim\omega^{-3}$ at $\omega\gg\omega_0$.
Also, it should be taken into account that this function gives a correct description of the resonance in the limit of its zero width only.
Given this, 
% Thus, from the theoretical viewpoint, 
the photon lineshape in the decay $J/\psi\to\gamma\eta_{\rm c}$ has the form
\begin{equation}
\label{ftwo}
\frac{d\Gamma(\omega)}{d\omega} \sim \omega^3 f(\omega) \textnormal{BW}(\omega),
\end{equation}
where the correction factor $f(\omega)$ is about one near the resonance and falls far from the resonance. 

Due to the $\omega^3$ factor and a fairly large $\eta_{\rm c}$ width, the photon lineshape in this decay is asymmetric,
and this is confirmed experimentally.
The Crystal Ball did not consider this issue in their publication, noting only that the $\omega^3$
factor was used in the fit of the spectrum in the convolution of the detector response function with the $\eta_{\rm c}$ 
Breit-Wigner resonance shape. 
However, because of the large background, such an asymmetry cannot be 
revealed using the data collected at the Crystal Ball.

The CLEO Collaboration used exclusive decay modes of the 
$\eta_{\rm c}$, that allows one to suppress background strongly. As a result, 
it was found that the photon lineshape of this transition 
is really asymmetric. The Breit-Wigner function alone, traditionally used to describe resonances, provides a poor fit to data. 
Its modification with the $\omega^3$ factor improves the fit around the peak, but gives a great tail at higher 
photon energies, as it was noted above.
To suppress this tail, CLEO used $|M|^2=\exp(-\frac{\omega^2}{8\beta^2})$ in their fit with $\beta=65$ MeV.
However, such a form of matrix element is valid for 
harmonic oscillator wave functions only.
Also, the value of $\beta$ used in the fit is too small for the 
charmonium system and
gives very fast fall of the matrix element with the photon energy increase. 
In addition, in their analysis CLEO did not consider 
interference effects, which may be not small for exclusive spectra.
%(see a discussion of interference effects in the next section).

When measuring the branching fraction ${\cal B}(J/\psi\to\gamma\eta_{\rm c})$, 
one should separate the events of 
$J/\psi\to\gamma\eta_{\rm c}$ decays from the background events. This requires 
either a knowledge of the photon lineshape or a background measurement
with sufficient accuracy. As a rule, the latter is a difficult task, 
especially for inclusive decays, because of the small
signal to background ratio. Therefore, to determine the number of signal events, during the data fitting one has to specify 
the explicit form of the resonance.
However, considering that exact $\omega$ dependence of the $f(\omega)$ factor in~(\ref{ftwo}) is unknown, we can conclude that the 
measurement of ${\cal B}(J/\psi\to\gamma\eta_{\rm c})$ will be inevitably model-dependent, until the photon lineshape will 
be measured or calculated theoretically with a sufficient accuracy. In this work we assume that the 
photon lineshape has the form~(\ref{ftwo}) wherein $f(\omega)$ is chosen 
under the assumption that the spectrum tail at photon energies 
$\omega-\omega_0>4\Gamma_{\eta_{\rm c}}$ can be neglected: at $\omega-\omega_0<2\Gamma_{\eta_{\rm c}}$ the factor $f(\omega)=1$,
  at $\omega-\omega_0>4\Gamma_{\eta_{\rm c}}$ the factor $f(\omega)=0$, and in the region 
$2\Gamma_{\eta_{\rm c}}<\omega-\omega_0<4\Gamma_{\eta_{\rm c}}$ the 
decay probability falls linearly.

\section{KEDR data}
The experiment was performed at the KEDR detector~\cite{KEDR} of the \mbox{VEPP-4M} collider~\cite{VEPP}. 
It operates at a peak luminosity of about $1.5\times 10^{30}$ ${\rm cm}^{-2}{\rm s}^{-1}$ near the $J/\psi$ resonance energy. 
The luminosity is measured using single Bremsstrahlung online and small-angle 
Bhabha scattering offline.
Two methods of a beam energy determination are used: a resonant depolarization with an accuracy of
$8\div30$ keV and an IR-light Compton backscattering with an accuracy 
of $\sim100$ keV~\cite{COMP}.

\begin{figure}
\includegraphics[width=\columnwidth]{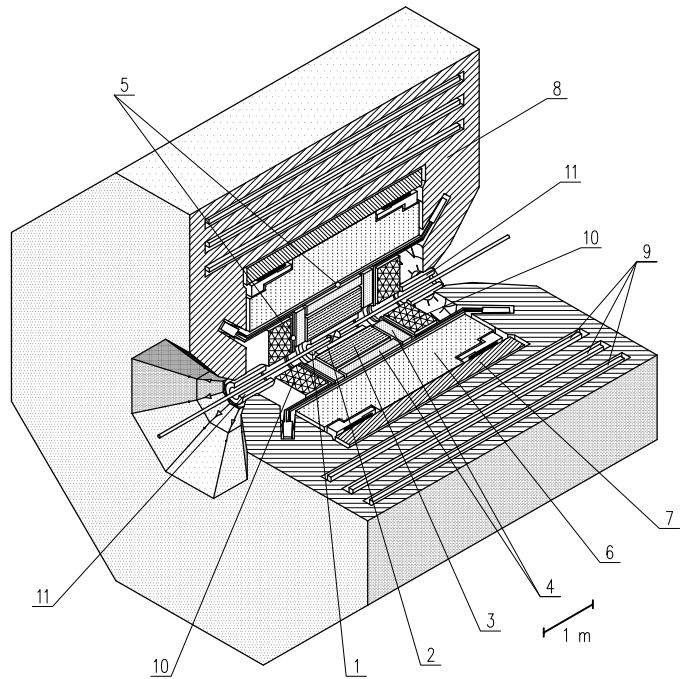}
\caption{The KEDR detector. 1~--- vacuum chamber, 2~--- vertex detector, 3~--- drift chamber, 4~--- threshold aerogel counters, 
5~--- time-of-flight counters, 6~--- liquid krypton calorimeter, 7~--- superconducting coil {(0.6~T)}, 8~--- magnet yoke,
9~--- muon tubes, 10~--- CsI-calorimeter, 11~--- compensating solenoid. \label{fig3}}
%%12~--- \mbox{VEPP-4M} quadrupole.
\end{figure}

The view of the KEDR detector is shown in Fig.~\ref{fig3}.  
Subsystems are listed in the figure. 
Detector includes a tracking system consisting of
a vertex detector and a drift chamber, a particle identification (PID)
system of aerogel Cherenkov counters and scintillation time-of-flight
counters, and an electromagnetic calorimeter based on liquid
krypton (in the barrel part) and CsI crystals (endcap part). The superconducting
solenoid provides a longitudinal magnetic field of
0.6 T. A muon system is installed inside the magnet yoke. The detector
also includes a high-resolution tagging system for studies of
two-photon processes. 

Charged tracks are reconstructed in the drift chamber (DC) and
vertex detector (VD). DC has a cylindrical shape, with a 1100 mm
length and an outer radius of 535 mm, and is filled with pure dimethyl
ether. DC cells form seven concentric layers: four axial layers
and three stereo layers to measure track coordinates along the
beam axis. The coordinate resolution averaged over drift length is
100 $\mu m$. VD is installed between the vacuum chamber and
DC and increases a solid angle accessible to the tracking system to
98$\%$. VD consists of 312 cylindrical drift tubes aligned in 6 layers.
It is filled with an Ar + $30\%$ CO$_2$ gas mixture and has a coordinate
resolution of 250 $\mu m$. The momentum resolution of the tracking
system is $\sigma_p/p = 2\%\oplus(4\%\times p [\rm GeV])$.

Scintillation counters of the time-of-flight system (TOF) are
used in a fast charged trigger and for identification of the charged
particles by their flight time. The TOF system consists of 32 plastic
scintillation counters in the barrel part and in each of the endcaps.
The flight time resolution is about 350 ps, which corresponds to
$\pi/K$ separation at the level of more than two standard deviations
for momenta up to 650 MeV.

Aerogel Cherenkov counters (ACC) are used for particle
identification in the momentum region not covered by the TOF
system and ionization measurements in DC. ACC uses aerogel
with a refractive index of 1.05 and wavelength shifters for light
collection. This allows one to identify $\pi$ and K mesons in the momentum
range of 0.6 to 1.5 GeV. The system includes 160
counters in the endcap and barrel parts, each arranged in two layers. 
During data taking only one layer of ACC was installed, and it
was not used because of insufficient efficiency.

The barrel part of the electromagnetic calorimeter is a liquid
krypton ionization detector. The calorimeter provides an energy
resolution of $3.0\%$ at the energy of 1.8 GeV and a spatial
resolution of $0.6\div1.0$ mm for charged particles and photons.
The endcap part of the calorimeter is based on 1536 CsI(Na) scintillation
crystals [18] with an energy resolution of $3.5\%$ at 1.8 GeV, and
a spatial resolution of 8 mm.

The muon system is used to identify muons by their flight
path in the dense medium of the magnetic yoke. It consists of
three layers of streamer tubes with $74\%$ solid angle coverage, the
total number of channels is 544. The average longitudinal resolution
is 3.5 cm, and the detection efficiency for most of the
covered angles is $99\%$.

The trigger of the KEDR detector has two levels: primary
(PT) and secondary (ST). Both PT and ST operate at the hardware
level. PT uses signals from TOF counters and both calorimeters as
inputs, its typical rate is $5\div10$ kHz. ST uses signals from VD, DC
and muon system in addition to the systems listed above, and the rate
is $50\div150$ Hz.

The analysis is based on a data sample of $(1.52\pm0.08)$ ${\rm pb}^{-1}$ collected at the $J/\psi$ peak and corresponding 
to about 6 million $J/\psi$ decays.
Photon selection was performed in two steps. At the first step 
multihadron decays of $J/\psi$ were selected. The 
following criteria suppressing 
backgrounds from cosmic rays, beam-gas interactions and Bhabha events, were applied: 
total energy in the calorimeters is greater than 0.8 GeV; 
at least four clusters with the energy greater than 30 MeV in 
the calorimeters are reconstructed; 
at least one central track in the drift chamber (DC) is reconstructed; 
there are no muon tubes activated in the third layer of the muon system. 
At the second step photons in these events were identified. 
A cluster in the liquid krypton calorimeter is considered as a photon if it is 
not associated with reconstructed tracks in the drift chamber 
and has no time-of-flight counters activated 
in front of it. According to a Monte Carlo simulation based on the GEANT3 package~\cite{GEANT}, the photon detection 
efficiency for the $J/\psi\to\gamma\eta_{\rm c}$ decay in the 
investigated energy range with the above criteria is nearly constant 
with sufficient accuracy.

The number of multihadron decays of $J/\psi$ selected at the first step 
of analysis is 
\begin{equation}
N_{\rm mh}^{\rm sel}=N_{\psi} {\cal B}_{\rm mh} \epsilon_{\rm mh}(1+b), 
\end{equation}
where ${\cal B}_{\rm mh}=87.7\%$~\cite{PDG} and $\epsilon_{\rm mh}$ are the 
branching fraction and selection 
efficiency for $J/\psi$ multihadron decays and $b$ is the fraction 
of nonresonant multihadron plus other background (mainly Bhabha) events that
passed selection criteria. 
The number of signal photons is 
\begin{equation}
N_{\rm sig}=N_{\psi}{\cal B}(J/\psi\to\gamma\eta_{\rm c})\epsilon'_{\rm mh}\epsilon_{\gamma},
\end{equation}
 where $\epsilon'_{\rm mh}$ is a selection efficiency 
for $\eta_{\rm c}$ multihadron decays and $\epsilon_\gamma$ is a photon selection efficiency. Hence
\begin{equation}
\label{fB}
{\cal B}(J/\psi\to\gamma\eta_{\rm c})={\cal B}_{\rm mh}\frac{N_{\rm sig}}{N_{\rm mh}^{sel}}\frac{\epsilon_{\rm mh}}{\epsilon'_{\rm mh}\epsilon_{\gamma}}(1+b). 
\end{equation}

According to the Monte Carlo simulation of $J/\psi$ decays using 
the generator~\cite{BESGEN}, based on the JETSET code~\cite{Jetset} and 
adopted by the BES Collaboration for charmonium decays, the 
selection efficiencies $\epsilon_{\rm mh}$ and $\epsilon'_{\rm mh}$ are close 
(87.9\% and 89.0\%, respectively). Many systematic errors appearing 
due to selection cuts substantially cancel in their ratio, 
so these efficiencies were taken from the simulation during the 
branching fraction calculation. The photon selection efficiency $\epsilon_{\gamma}$ was determined by imposition of 
MC photons on the multihadron events selected 
in the experimental $J/\psi$ decays. A small correction was applied to take into account a difference between the selection efficiency of 
photons, imposed on MC $J/\psi\to\gamma\eta_{\rm c}$ decays and $J/\psi$ multihadron decays. This difference was taken as an 
estimate of the systematic error for $\epsilon_{\gamma}$.
The fraction $b$ was determined from the data sample 
collected at the energy of 10 MeV below the $J/\psi$ resonance  and 
is equal to $(3.6\pm0.5)\%$.

\begin{figure}
\includegraphics[width=\columnwidth]{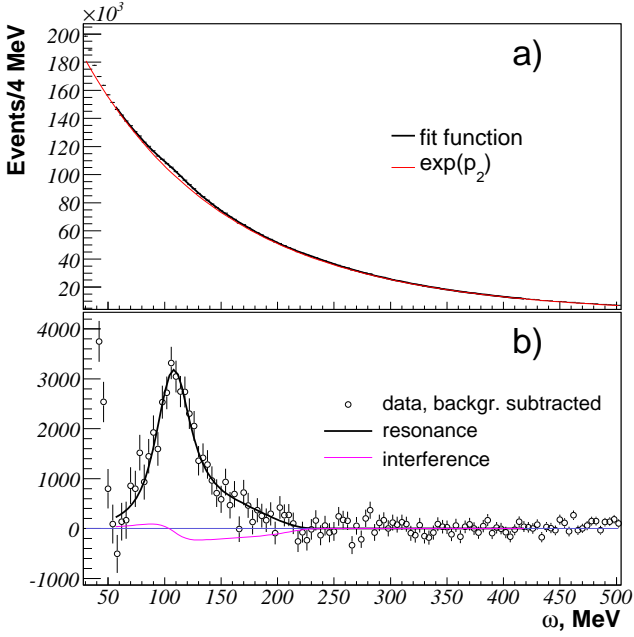}
\caption{a) The fit of the inclusive photon spectrum 
in the energy range 55-420 MeV. 
b) The photon spectrum after background subtraction. \label{fig5}}
\end{figure}

In Fig.~\ref{fig5}a,b the inclusive photon spectrum and its fit are shown.
The spectrum was fit with a sum of the signal having the shape~(\ref{ftwo}), 
convolved with the calorimeter response function, 
and background. The calorimeter response function was approximated 
with a logarithmic normal distribution~\cite{PROT} 
with $\sigma_{\rm E}=6.7$ MeV at 110 MeV and asymmetry $a=-0.26$. 

The background has the following shape: 
\begin{equation}
\label{f3}
dN/d\omega=\exp(p_2(\omega))+c\times \textnormal{MIP}(\omega), 
\end{equation}
where $p_2(\omega)$ is a second-order 
polynomial and MIP$(\omega)$ is the spectrum of charged particles.
The first term in the expression~(\ref{f3}) well describes background 
in the photon energy range of $50\div450$ MeV. 
These photons arise mainly from $\pi^0$ decays, other processes give a 
small contribution.
At lower photon energies background is not described by such a simple 
form because of the additional 
significant contribution from neutral clusters 
appearing due to nuclear interactions of hadrons in the calorimeter. 
Due to inefficiencies of track reconstruction and TOF counters, a 
small part of charged particles is misidentified as neutrals,
so the scaled histogram of charged particles was added to the background 
function during the fit of the photon spectrum.
In the fit,  parameters of the polynomial and $c$ coefficient were varied 
freely.
The number of signal photons determined from the fit is equal to 
$N_{\rm sig}=(45.4\pm2.9)\times 10^3$, while the number of the multihadron 
events selected is $N_{\rm mh}^{\rm sel}=4.70\times 10^6$.

The fit gives the following values of the $\eta_{\rm c}$ mass, width and 
branching fraction of $J/\psi\to\gamma\eta_{\rm c}$ decay: 
$M_{\eta_{\rm c}} = (2982.6 \pm 1.7)$ MeV/$c^2$, 
$\Gamma_{\eta_{\rm c}} = (27.2 \pm 3.1)$ MeV and
 ${\cal B}(J/\psi\to\gamma\eta_{\rm c}) = (3.40\pm0.33)\%$.
The mass and width values are determined from the spectrum shape in the 
region of the resonance peak, 
thus for them the model uncertainty related to lineshape is small. 
At the same time 
this uncertainty for the branching fraction is mainly determined by 
the tail of the spectrum and is
much larger. The fit of the spectrum using the lineshape~(\ref{ftwo}) 
with $f(\omega)\equiv 1$ gives
${\cal B}(J/\psi\to\gamma\eta_{\rm c}) = (10.3\pm0.6)\%$, i.e. the decay rate 
can, in principle, be determined just by the tail. Thus, the large model 
uncertainty for the branching fraction makes its 
measurement hardly meaningful.

However, it is possible to define another quantity, which characterizes
the decay rate and is less model-dependent. To do that, let us write the 
photon spectrum of decay in the form
\begin{equation}
  \begin{aligned}
\frac{d\Gamma}{d\omega}&=\frac{d\Gamma}{d\omega}(\omega_0) \left(\frac{\omega}{\omega_0}\right)^3 \frac{f(\omega)}{f(\omega_0)} \frac{BW(\omega)}{BW(\omega_0)}=\\
                       &=\Gamma^0_{\gamma\eta_{\rm c}}\left(\frac{\omega}{\omega_0}\right)^3 \frac{f(\omega)}{f(\omega_0)} BW(\omega), 
  \end{aligned}
\end{equation}
where
\begin{equation}
\Gamma^0_{\gamma\eta_{\rm c}}=\frac{1}{BW(\omega_0)} \frac{d\Gamma}{d\omega}(\omega_0)=\frac{\Gamma_{\gamma\eta_{\rm c}}}{f_{\rm cor}}, 
\end{equation}
\begin{equation}
\label{ffcor}
f_{\rm cor}=\int_0^{M_{\psi}/2} \left(\frac{\omega}{\omega_0}\right)^3 \frac{f(\omega)}{f(\omega_0)} BW(\omega) d\omega.
\end{equation}
The resonance height in the fit weakly depends on the lineshape chosen, 
because due to the $\omega^3$ factor the spectrum quickly tends to zero 
to the left of the resonance. Thus, the measured $\Gamma^0_{\gamma\eta_{\rm c}}$ value 
has small model uncertainty. Besides, if the resonance width tends to zero, 
the factor $f_{\rm cor}$ tends to unity, 
i.e. $\Gamma^0_{\gamma\eta_{\rm c}}$ is the partial decay width in the case of 
a narrow resonance, and can be directly compared to theoretical 
calculations~\cite{SHIF,KHOD,BEIL,BRAM,DUDE,DONALD,Becirevic,Pineda}. 
Thus, this quantity has clear 
physical meaning and can be used  as a characteristic of the decay rate. 
For our lineshape model the factor $f_{\rm cor}$ is about 1.12 and 
$\Gamma^0_{\gamma\eta_{\rm c}}=2.86\pm0.28$ keV.

A statistical error of the $\eta_{\rm c}$ width obtained in the fit 
is much larger than the accuracy of its world average of 
$(29.7\pm1.0)$ MeV, therefore the final values for mass and 
$\Gamma^0_{\gamma\eta_{\rm c}}$ are obtained from the fit with fixed 
$\Gamma_{\eta_{\rm c}} = 29.7$ MeV:
$M_{\eta_{\rm c}} = (2983.5 \pm 1.4)$ MeV/$c^2$, 
$\Gamma^0_{\gamma\eta_{\rm c}}=2.98\pm0.18$ keV.
A systematic error related to the uncertainty of the $\eta_{\rm c}$ width is
estimated varying this value in the fit by 1.0 MeV.

The above results were obtained without taking into account 
interference effects.
However, decays $J/\psi\to\gamma\eta_{\rm c}, \eta_{\rm c}\to X$ 
can interfere with other radiative decays of $J/\psi$ into the same 
final multihadron state $X$.
At first glance, for the inclusive spectrum these effects should be small 
due to a lot (many dozens) of  $\eta_{\rm c}$ decay channels
and different relative phases of interference. 
However, recently the BESIII Collaboration published~\cite{BESetacp} 
results of a measurement of the $\eta_{\rm c}$ mass and width analyzing six 
exclusive decay modes of $\psi(2S)\to\gamma\eta_{\rm c}$ decay, 
where it was found that the phases of interference 
with nonresonant background are close to each other for all decay modes. 
If the same holds for 
$J/\psi\to\gamma\eta_{\rm c}$ decays, then the interference effects for 
the inclusive spectrum may be not small and should be  also taken into account.

First of all, note that the $J/\psi\to\gamma\eta_{\rm c}\to\gamma X$ 
decay amplitude can interfere with the amplitude of 
the $J/\psi\to \gamma gg \to\gamma X$ decay.
Since the $\eta_{\rm c}$ meson also decays mainly through two gluons, 
the lower-order Feynman diagrams for these processes are the same.
Therefore it can be assumed that in these decays the relative 
interference phases are close for all decay channels
(if the quantum numbers of the final systems are the same).
Second, processes $J/\psi\to q\bar q \to (\gamma)X$, 
$J/\psi\to ggg \to (\gamma)X$, when one of the final hadrons radiates 
an additional photon (FSR), should be  also taken into account. 
However, in this case the diagrams of these processes are different, 
therefore it is reasonable to assume that the
relative phases are different as well.

According to this, the inclusive photon spectrum in the
$J/\psi\to\gamma\eta_{\rm c}$ decay taking into account interference with $J/\psi\to \gamma gg \to\gamma X$ decays 
can be written in the form\begin{equation}
\label{fspecI0}
  \begin{aligned}
&\frac{d\Gamma(\omega)}{d\omega} \sim \sum\limits_k\left|S_k + N_k\right|^2= \\
& S^2+N^2+2SN\cos(\gamma-\phi)\sum\limits_k\frac{|S_k|}{S}\frac{|N_k|}{N},
  \end{aligned}
\end{equation}
%$x=\frac{\omega}{\omega_0}$,
where $S_k=x^{3/2}f(\omega)^{1/2}\frac{\sqrt{s\Gamma_k}}{s-M_{\eta_{\rm c}}^2+i\sqrt{s}\Gamma_{\eta_{\rm c}}}$ are resonant amplitudes, 
$N_k$ - nonresonant amplitudes of the k-th channel of $J/\psi$ decays through $\gamma gg$, 
$x=\frac{\omega}{\omega_0}$, $s=M_{\psi}^2-2\omega M_{\psi}$, $S=\sqrt{\sum\nolimits_k\left|S_k\right|^2}$, 
$N=\sqrt{\sum_k\left|N_k\right|^2}$, $\gamma$ and $\phi$ are resonant and nonresonant phase, respectively. 
Partial widths for these decays are known for few decay channels, so we can only estimate an upper bound of this 
interference contribution, replacing the 
sum $\sum_k\frac{|S_k|}{S}\frac{|N_k|}{N}$ 
in the expression with unity. After that~(\ref{fspecI0}) takes the form which is analogous 
to interference in the single decay channel:
\begin{equation}
\label{fspecI}
  \begin{aligned}
\frac{d\Gamma(\omega)}{d\omega} \sim 
\left|Se^{i\gamma} + x^{1/2}N(\omega_0)e^{i\phi} \right|^2,
  \end{aligned}
\end{equation}
where for a nonresonant term the explicit energy dependence according 
to~\cite{KW} is specified.

To estimate the $N(\omega)$ magnitude, MC simulation of 
$J/\psi$ decays using the generator~\cite{BESGEN} was performed.
The generator poorly reproduces the experimental 
photon spectrum of $J/\psi \to \gamma gg$ decays, 
thus $N(\omega_0)$ from the simulation has been corrected using
data for this process~\cite{BESSON}. An additional correction was made 
assuming that only a fraction of the $J/\psi \to \gamma gg$  decay amplitude
with the same quantum numbers of the $gg$-system and $\eta_{\rm c}$ interferes. 
The probability for the $gg$-system to have $J^P=0^-$
in this decay was calculated in the lowest order in~\cite{BILLOIRE} 
and equals 0.3 for small $\omega$. 
With these corrections we estimate the $N(\omega_0)$ value as 
$(4.6\pm2.2)\%$ of $S(\omega_0)$.
The FSR contribution to interference was also estimated with the help 
of additional simulation in which the final state radiation was modeled 
using the PHOTOS~\cite{PHOTOS} package,
and phases of different decay channels were generated randomly. Its value 
was found to be small compared to the $\gamma gg$ contribution.

The fit taking into account interference according to the 
expression~(\ref{fspecI}), with 
$N(\omega_0)$ fixed to $4.6\%$ of $S(\omega_0)$ and phase $\phi$ varied 
freely, gives the following values 
of $\eta_{\rm c}$ mass, width and decay rate:  
$M_{\eta_{\rm c}} = (2981.6 \pm 1.9)$ MeV/$c^2$, 
$\Gamma_{\eta_{\rm c}} = (29.9 \pm 3.4)$ MeV, 
$\Gamma^0_{\gamma\eta_{\rm c}}=2.82\pm0.37$ keV.
In Fig.~\ref{fig5}b the interference contribution in this case is shown. 
The value of the obtained phase $\phi=(-4\pm54)^{\circ}$ is close to zero, 
so the values of parameters for the second solution are 
almost the same. 
The magnitude of $N(\omega_0)$ is quite small, therefore interference changes 
the measured values only slightly. 
These shifts are considered as systematic uncertainties due to 
interference effects.

The main resulting systematic uncertainties are shown in Table~\ref{ta3}.
\begin{table*}
\begin{center}
\caption{Systematic uncertainties. \label{ta3}}
\vspace*{\baselineskip}
\begin{tabular}{@{}lcccc@{}} \hline
Systematic error & $M_{\eta_{\rm c}}$, MeV/$c^2$ & $\Gamma_{\eta_{\rm c}}$, MeV  & $\Gamma^0_{\gamma\eta_{\rm c}}$, keV \\
\hline
 Background subtraction & 0.8 & 1.4  & 0.11  \\
 Calorimeter response function & 2.2 & 0.8 & 0.07\\
 Lineshape & 0.7 & 2.8 & 0.05  \\
 $\eta_{\rm c}$ width & 0.3 & - & 0.06  \\
 Interference effects & -2.1 & +2.3 & -0.18 \\
 Photon selection efficiency & - & - & 0.16  \\
 $J/\psi$ width & - & - & 0.09  \\
\hline
\end{tabular} 
\end{center}
\end{table*}
To estimate the systematic uncertainty related to the background 
subtraction, we varied the range of the fit,
changed the order of the polynomial in the first term of~(\ref{f3}) 
from the second to third, and fitted the spectrum without taking 
into account time-of-flight counters. 
The systematic errors for the ${\eta_{\rm c}}$ mass, 
width and $\Gamma^0_{\gamma\eta_{\rm c}}$, 
appearing due to a poorly known photon lineshape were estimated 
by changing the low energy cut-off parameter 
$2\Gamma_{\eta_{\rm c}}$ to $1.5\Gamma_{\eta_{\rm c}}$, 
and taking $f(\omega)\equiv 1$. 
The calibration of the photon energy scale was performed using 
$\pi^0\to 2\gamma$ decays. Within $1.5\%$ it agrees with the calibrations 
made with a data sample  collected at the $\psi(2S)$ peak, using 
$\psi(2S)\to\gamma\chi_{c1}$, $\psi(2S)\to\gamma\chi_{c2}$ transitions. 
No scale shift was observed for different seasons of collecting data. 
The systematic error related to the shape of the 
calorimeter response function was estimated by varying parameters 
$\sigma_E$ and $a$ of a logarithmic normal distribution. 
Shifts of measured values due to interference effects in the table 
are given with signs.

\section{Results and conclusions}
A new direct measurement of $J/\psi\to\gamma\eta_{\rm c}$ decay was performed. 
We measured the $\eta_{\rm c}$ mass, width and decay rate 
$\Gamma^0_{\gamma\eta_{\rm c}}$ of
the $J/\psi\to\gamma\eta_{\rm c}$ decay.
These parameters are sensitive to the lineshape of the photon spectrum 
in this decay and it was taken into account during analysis. 

Our results for the $\eta_{\rm c}$ mass and width are
\begin{center}
$M_{\eta_{\rm c}} = 2983.5 \pm 1.4 \phantom{|}^{+1.6}_{-3.6}$ MeV/$c^2$, 
\end{center}
\begin{center}
$\Gamma_{\eta_{\rm c}} = 27.2 \pm 3.1 \phantom{|}^{+5.4}_{-2.6}$ MeV. 
\end{center}
These parameters were earlier measured in $J/\psi$ and 
$B$ meson decays as well as in $\gamma\gamma$ and $p\bar p$ collisions. 
Measurements of Crystal Ball~\cite{GAIS}, MARK3~\cite{MRKIII}, 
BES~\cite{BES00F,BES03}, and KEDR were performed 
using radiative decays of the $J/\psi$ resonance, therefore 
a mass shift due to an asymmetric lineshape should be taken 
into account. Crystal Ball and KEDR made 
such a correction in their experiments, whereas MARK3 and BES did not. 
Therefore  we believe that MARK3 and BES results on the $\eta_{\rm c}$ 
mass should be corrected by approximately 4 MeV towards higher values 
due to this effect. Interference effects for exclusive 
decays may give not small shifts and should be also analyzed. 

\begin{figure}
\includegraphics[width=\columnwidth]{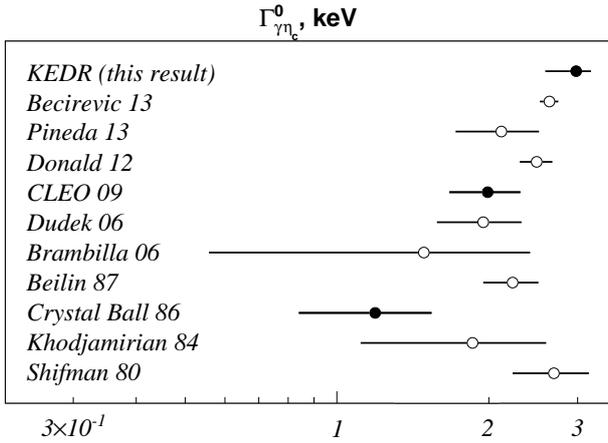}
\caption{Results of measurements (close circles) and theoretical predictions (open circles) on 
$\Gamma^0_{\gamma\eta_{\rm c}}$. 
\label{fig9}}
\end{figure}

Our result on the decay rate is
\begin{center}
$\Gamma^0_{\gamma\eta_{\rm c}}=2.98\pm0.18 \phantom{|}^{+0.15}_{-0.33}$ keV.
\end{center}
In Fig.~\ref{fig9} this result is compared with the 
Crystal Ball and CLEO measurements as well as with theoretical predictions. 
The Crystal Ball and CLEO results on $\Gamma^0_{\gamma\eta_{\rm c}}$
were evaluated using their measured branching fractions from the formula 
\begin{equation}
\Gamma^0_{\gamma\eta_{\rm c}}=\frac{{\cal B}(J/\psi\to\gamma\eta_{\rm c})\Gamma_{\psi}}{f_{\rm cor}},
\end{equation}
where $\Gamma_{\psi}=92.9\pm2.8$ keV is the $J/\psi$ width. Factors 
$f_{\rm cor}$ were calculated from formula~(\ref{ffcor}).
For Crystal Ball the function $f(\omega)\equiv 1$ was taken and 
integration was made from 40 to 165 MeV, 
which corresponds to the range of their spectrum fit. For CLEO the 
function $f(\omega)=\exp(-\frac{\omega^2}{8\beta^2})$ and 
${\cal B}(J/\psi\to\gamma\eta_{\rm c})=(2.06\pm0.32)\%$~\cite{PDG} were used.
The resulting values of $f_{\rm cor}$ equal 0.96 and 0.99, 
and partial widths are $1.23\pm0.35$ keV and $1.93\pm0.31$ keV, respectively.
Our decay rate value is significantly higher compared to those experimental 
results,
but is well consistent with the latest lattice QCD prediction~\cite{Becirevic}: 
$\Gamma_{\gamma\eta_{\rm c}}=(2.64\pm0.11)$ keV.

The authors are grateful to N. Brambilla, V.L. Chernyak,  A.I. Milstein
and A. Vairo for useful discussions. 
This work was supported by the Ministry of Education and
Science of the Russian Federation, grant No. 14.518.11.7003, 
RFBR grants 08-02-00258, 10-02-00695,
11-02-00112, 11-02-00558, 12-02-00023, 12-02-01032, 12-02-91341, RF 
Presidential Grants for Sc. Sch. NSh-5655.2008.2, NSh-5320.2012.2 and NSh-2479.2014.2
as well as the DFG grant GZ: HA 1457/9-1.



%% References
%%   ~\cite{key}         ==>>  [#]


%\bibliographystyle{elsarticle-num}






% \end{thebibliography}

\end{document}